\documentstyle[sprocl]{article}

\input{psfig}

\def\be{\begin{equation}}
\def\ee{\end{equation}}
\def\bea{\begin{eqnarray}}
\def\eea{\end{eqnarray}}

\makeatletter
\newbox\slashbox \setbox\slashbox=\hbox{\large$/$}
\def\pslash#1{\setbox\@tempboxa=\hbox{$#1$}
  \@tempdima=0.5\wd\slashbox \advance\@tempdima 0.5\wd\@tempboxa
  \copy\slashbox \kern-\@tempdima \box\@tempboxa}
\def\FMSlash{\protect\pslash}

\def\@cite#1#2{{#1\if@tempswa , #2\fi}}
\def\@citex[#1]#2{%
\if@filesw\immediate\write\@auxout{\string\citation{#2}}\fi
\leavevmode\unskip$^{\,\scriptstyle\@cite{\@collapse{#2}}{#1}}$}
\def\CITE{\@ifnextchar[{\@tempswatrue\@CITEX}{\@tempswafalse\@CITEX[]}}
\let\onlinecite\CITE
\def\@CITEX[#1]#2{%
\if@filesw\immediate\write\@auxout{\string\citation{#2}}\fi
\leavevmode\unskip\ \@cite{\@collapse{#2}}{#1}}
\def\@collapse#1{{%
\let\@temp\relax
\@tempcntb\@MM
\def\@citea{}%
\@for \@citeb:=#1\do{%
\@ifundefined{b@\@citeb}%
{\@temp\@citea{\bf ?}%
\@tempcntb\@MM\let\@temp\relax
\@warning{Citation `\@citeb ' on page \thepage\space undefined}}%
{\@tempcnta\@tempcntb \advance\@tempcnta\@ne
\edef\MyTemp{\csname b@\@citeb\endcsname}%
\def\@tempa{\@temptokena=\bgroup}%
\afterassignment\@tempa %
\@tempcntb=0\MyTemp\relax}%
\ifnum\@tempcntb=0\relax%
\@tempcntb=\@MM
\@citea\MyTemp
\let\@temp = \relax
\else %
\edef\@tempd{\number\@tempcntb}%
\ifnum\@tempcnta=\@tempcntb %
\ifx\@temp\relax %
\edef\@temp{\@citea\@tempd}%
\else
\edef\@temp{\hbox{--}\@tempd}%
\fi
\else %
\@temp\@citea\@tempd
\let\@temp\relax
\fi
\fi
}%
\def\@citea{, }%
}%
\@temp %
}}%

\makeatother

\begin{document}

\title{UNIVERSALITY AND CHAOS IN QUANTUM FIELD THEORIES}

\author{B.A. BERG}
\address{Department of Physics, The Florida State University,
  Tallahassee, FL 32306} 
\author{E. BITTNER, H. MARKUM, R. PULLIRSCH}
\address{Institut f\"ur Kernphysik, Technische Universit\"at Wien, \\
                A-1040 Vienna, Austria}
\author{M.-P. LOMBARDO}
\address{Istituto Nazionale di Fisica Nucleare,
         Laboratori Nazionali del Gran Sasso, \\  I-67010 Assergi, Italy}
\author{T. WETTIG}
\address{Department of Physics, Yale University, New Haven, CT
  06520-8120 and\\
  RIKEN BNL Research Center, Upton, NY 11973-5000}

\maketitle

\abstracts{
  We investigate the eigenvalue spectrum of the staggered Dirac matrix
  in SU(3) gauge theory and in full QCD as well as in quenched U(1)
  theory on various lattice sizes.  As a measure of the fluctuation
  properties of the eigenvalues, we consider the nearest-neighbor spacing
  distribution, $P(s)$.  
  We further study two-color QCD at nonzero chemical potential, $\mu$, by
  constructing the spacing distribution of adjacent eigenvalues in the
  complex plane.  We find that in all regions of their phase diagrams,
  compact lattice gauge theories have bulk spectral correlations given
  by random matrix theory, which is an indication for quantum chaos.
  In the confinement phase, the low-lying Dirac spectrum of these quantum field
  theories is well described by random matrix theory, exhibiting universal
  behavior.
} 

\section{Bulk of the Spectrum}

The properties of the eigenvalues of the Dirac operator are of great
interest for the universality of certain features of QCD and QED.  On
the one hand, the accumulation of small eigenvalues is, via the
Banks-Casher formula,\cite{Bank80} related to the spontaneous
breaking of chiral symmetry. On the other hand, the fluctuation
properties of the eigenvalues in the bulk of the spectrum have also
attracted attention.  It was shown in Ref.~\onlinecite{Hala95} that on
the scale of the mean level spacing they are described by random
matrix theory (RMT). For example, the nearest-neighbor spacing
distribution $P(s)$, i.e., the distribution of spacings $s$ between
adjacent eigenvalues on the unfolded scale, agrees with the Wigner
surmise of RMT.  According to the Bohigas-Giannoni-Schmit
conjecture,\cite{Bohi84} quantum systems whose classical counterparts are
chaotic have a nearest-neighbor spacing distribution given by RMT
whereas systems whose classical counterparts are integrable obey a
Poisson distribution, $P_{\rm P}(s)=e^{-s}$.  Therefore, the specific
form of $P(s)$ is often taken as a criterion for the presence or
absence of ``quantum chaos''.

In RMT, one has to distinguish several universality classes which are
determined by the symmetries of the system.  For the case of the QCD
Dirac operator, this classification was done in
Ref.~\onlinecite{Verb94}.  Depending on the number of colors and the
representation of the quarks, the Dirac operator is described by one
of the three chiral ensembles of RMT.  As far as the fluctuation
properties in the bulk of the spectrum are concerned, the predictions
of the chiral ensembles are identical to those of the ordinary
ensembles.\cite{Fox64}  In Ref.~\onlinecite{Hala95}, the Dirac matrix
was studied for color-SU(2) using both staggered and Wilson fermions which
correspond to the chiral symplectic (chSE) and orthogonal (chOE) ensemble,
respectively.  Here,\cite{Pull98} we study SU(3) with staggered
fermions which corresponds to the chiral unitary ensemble (chUE). The
RMT result for the nearest-neighbor spacing distribution can be
expressed in terms of so-called prolate spheroidal functions, see
Ref.~\onlinecite{Meht91}.  A very good approximation to $P(s)$ is
provided by the Wigner surmise for the unitary ensemble,
\begin{equation} \label{wigner}
  P_{\rm W}(s)=\frac{32}{\pi^2}s^2e^{-4s^2/\pi} \:.
\end{equation}

We generated gauge field configurations using the standard Wilson
plaquette action for SU(3) with and without dynamical fermions in the
Kogut-Susskind prescription. We have worked on a $6^3\times 4$ lattice
with various values of the inverse gauge coupling $\beta=6/g^2$ both
in the confinement and deconfinement phase.  We typically produced 10
independent equilibrium configurations for each $\beta$.  Because of
the spectral ergodicity property of RMT one can replace ensemble
averages by spectral averages if one is only interested in bulk
properties.

The Dirac operator, $\FMSlash{D}=\FMSlash{\partial}+ig\FMSlash{A}$, is
anti-Hermitian so that the eigenvalues $\lambda_n$ of $i\FMSlash{D}$
are real.  Because of $\{\FMSlash{D},\gamma_5\}=0$ the nonzero
$\lambda_n$ occur in pairs of opposite sign.  All spectra were checked
against the analytical sum rules $\sum_{n} \lambda_n = 0$ and
$\sum_{\lambda_n>0} \lambda_n^2 = 3V$, where V is the lattice volume.
To construct the nearest-neighbor spacing distribution from the
eigenvalues, one first has to ``unfold'' the spectra.\cite{Meht91}

\begin{figure}
\begin{center}
\begin{tabular}{ccccc}
  & {Confinement $\beta=5.2$}  & \hspace*{8mm}   & &
  {Deconfinement $\beta=5.4$} \\
  \vspace*{0mm}
  & $ma=0.05$ &&& $ma=0.05$ \\[2mm]
  \multicolumn{2}{c}{\epsfxsize=4.75cm\epsffile{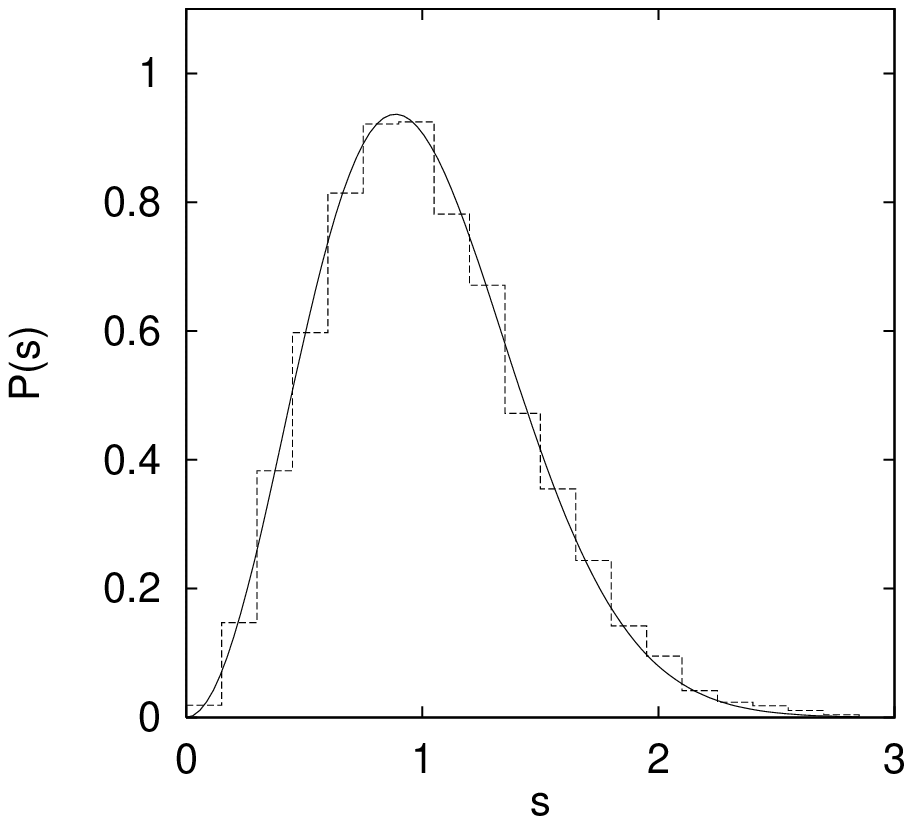}} &&
  \multicolumn{2}{c}{\epsfxsize=4.75cm\epsffile{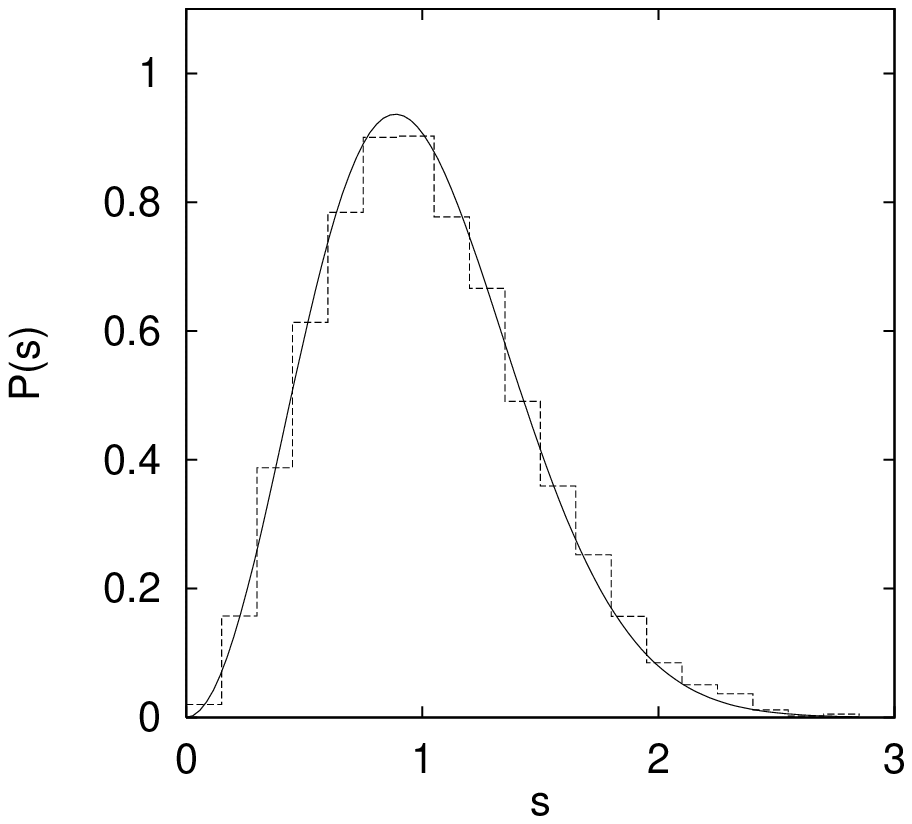}}
\end{tabular}
\end{center}
\vspace*{-3mm}
\caption{Nearest-neighbor spacing distribution $P(s)$ on a $6^3 
  \times 4$ lattice in full QCD (histograms) compared with the
  random matrix result (solid lines). There are no changes in $P(s)$
  across the deconfinement phase transition.}
\vspace*{-3mm}
\label{fintemp}
\end{figure}

Figure~\ref{fintemp} compares $P(s)$ of full QCD with $N_f = 3$
flavors and quark mass $ma=0.05$ to the RMT result.  In the confinement
as well as in the deconfinement phase we observe agreement with RMT up
to very high $\beta$ (not shown).  The observation that $P(s)$ is not
influenced by the presence of dynamical quarks is
expected from the results of Ref.~\onlinecite{Fox64}, which
apply to the case of massless quarks. Our
results, and those of Ref.~\onlinecite{Hala95}, indicate that massive
dynamical quarks do not affect $P(s)$ either.

\begin{figure}[h]
  \centerline{\psfig{figure=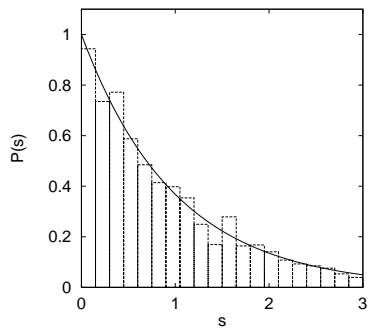,width=4.75cm}}
  \caption{Nearest-neighbor spacing distribution $P(s)$ for the free
            Dirac operator on a $53\times 47\times 43\times 41$ lattice
            compared with a Poisson distribution, $e^{-s}$.}
  \label{free}
\end{figure}

No signs for a transition to Poisson regularity are found. The
deconfinement phase transition does not seem to coincide with a
transition in the spacing distribution. For very large values of
$\beta$ far into the deconfinement region, the eigenvalues
start to approach the degenerate eigenvalues of the free theory, given
by $\lambda^2=\sum_{\mu=1}^4 \sin^2(2\pi n_\mu/L_\mu)/a^2$, where $a$
is the lattice constant, $L_{\mu}$ is the number of lattice sites in
the $\mu$-direction, and $n_\mu=0,\ldots,L_\mu-1$.  In this case, the
spacing distribution is neither Wigner nor Poisson.
It is possible to lift the degeneracies of the free
eigenvalues using an asymmetric lattice where $L_x$, $L_y$, etc. are
relative primes and, for large lattices, the distribution 
is then Poisson, $P_{\rm P}(s)=e^{-s}$, see Fig.~\ref{free}.

\begin{figure*}[t]
  \centerline{\psfig{figure=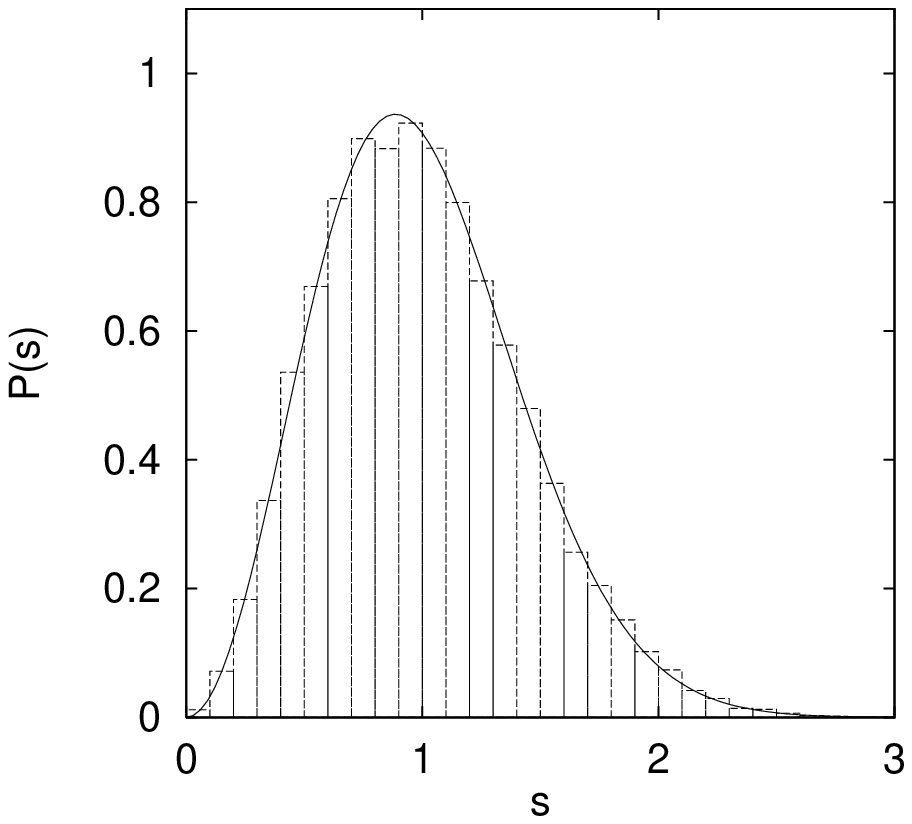,width=5cm}\hspace*{10mm}
    \psfig{figure=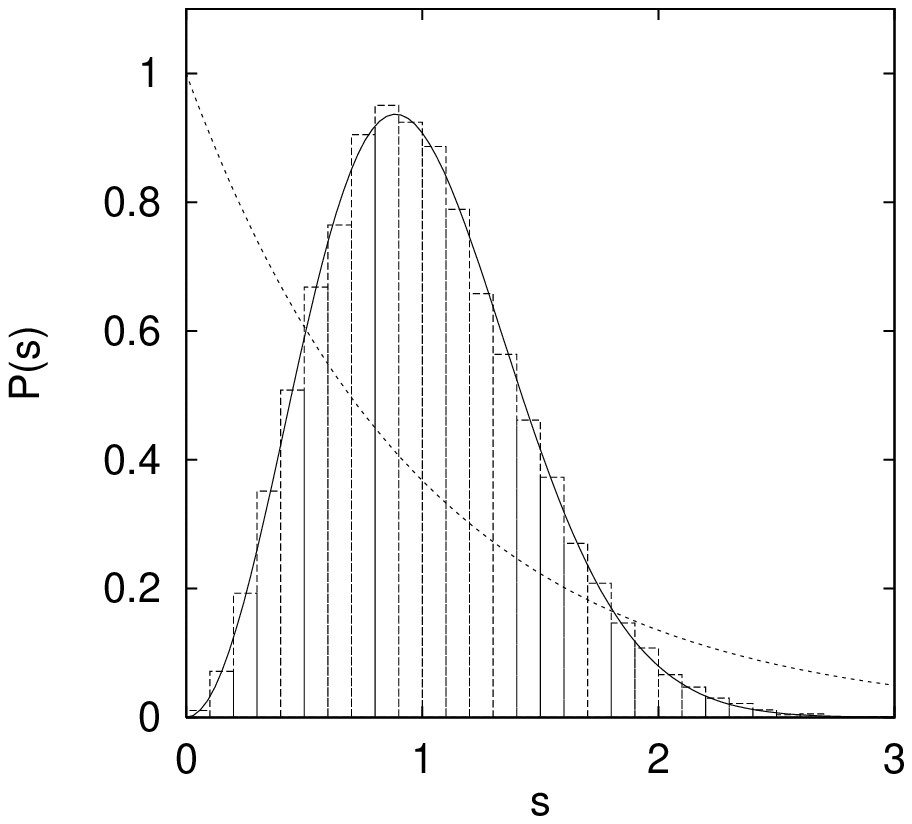,width=5cm}}
  \caption{Nearest-neighbor spacing distribution $P(s)$ for U(1) gauge
    theory on an $8^3\times 6$ lattice in the confined phase (left)
    and in the Coulomb phase (right). The theoretical curves are the chUE
    result, Eq.~(\ref{wigner}), and the Poisson distribution, $P_{\rm
      P}(s)=\exp(-s)$.}
  \label{f02}
\end{figure*}

We have also investigated the staggered Dirac spectrum of 4d U(1)
gauge theory which corresponds to the chUE of RMT but had
not been studied before in this context.  At $\beta_c \approx 1.01$
U(1) gauge theory undergoes a phase transition between a confinement
phase with mass gap and monopole excitations for $\beta < \beta_c$ and
the Coulomb phase which exhibits a massless photon for $\beta >
\beta_c$.\cite{BePa84} As for SU(2) and SU(3) gauge groups, we
expect the confined phase to be described by RMT, whereas free
fermions are known to yield the Poisson distribution (see
Fig.~\ref{free}). The question arose whether the Coulomb phase would be
described by RMT or by the Poisson distribution.\cite{BeMaPu99}  The
nearest-neighbor spacing distributions for an $8^3\times 6$ lattice at
$\beta=0.9$ (confined phase) and at $\beta=1.1$ (Coulomb phase),
averaged over 20 independent configurations, are depicted in
Fig.~\ref{f02}. Both are well described by the chUE of RMT.

Physical systems which are described by non-Hermitian operators have
attracted a lot of attention recently, among others QCD at nonzero
chemical potential $\mu$.\cite{Step96}
A formulation of the QCD Dirac
operator at $\mu\ne0$ on the lattice in the staggered scheme is
given by \cite{Hase83}
\begin{eqnarray}
  \label{Dirac}
  M_{x,y}(U,\mu) & = &
  \frac{1}{2a} \sum\limits_{\nu=\hat{x},\hat{y},\hat{z}}
  \left[U_{\nu}(x)\eta_{\nu}(x)\delta_{y,x\!+\!\nu}-{\rm h.c.}\right]
  \nonumber\\
  & + &
   \frac{1}{2a}\left[U_{\hat{t}}(x)\eta_{\hat{t}}(x)e^{\mu}
    \delta_{y,x\!+\!\hat{t}}
    -U_{\hat{t}}^{\dagger}(y)\eta_{\hat{t}}(y)
    e^{-\mu}\delta_{y,x\!-\!\hat{t}}\right]
\end{eqnarray}
with the link variables $U$ and the staggered phases $\eta$.
For $\mu>0$, the Dirac operator
loses its Hermiticity properties so that its eigenvalues become
complex. The aim of the present analysis is to investigate whether
non-Hermitian RMT is able to describe the fluctuation properties of
the complex eigenvalues of the QCD Dirac operator.  The eigenvalues
are generated on the lattice for various values of $\mu$.  We apply a
two-dimensional unfolding procedure\cite{Mark99} to separate the
average eigenvalue 
density from the fluctuations and construct the nearest-neighbor
spacing distribution, $P(s)$, of adjacent eigenvalues in the complex
plane. Adjacent eigenvalues are defined to be the pairs for which the
Euclidean distance in the complex plane is smallest.  
The data are then compared to analytical predictions of the Ginibre
ensemble \cite{Gini65} of non-Hermitian RMT, which describes the
situation where the real and imaginary parts of the strongly
correlated eigenvalues have approximately the same average magnitude.
In the Ginibre ensemble, the average spectral density is already
constant inside a circle and zero outside.  In this
case, unfolding is not necessary, and $P(s)$ is given by \cite{Grob88}
\begin{equation} \label{Ginibre}
  P_{\rm G}(s)  =  c\, p(cs)\:, ~~p(s) = 
  2s\lim_{N\to\infty}\left[\prod_{n=1}^{N-1}e_n(s^2)\,e^{-s^2}
  \right] \sum_{n=1}^{N-1}\frac{s^{2n}}{n!e_n(s^2)}\:,
\end{equation}
where $e_n(x)=\sum_{m=0}^n x^m/m!$ and $c=\int_0^\infty ds \, s \,
p(s)=1.1429...$. 
For uncorrelated eigenvalues in the
complex plane, the Poisson distribution becomes \cite{Grob88}
\begin{equation}
  \label{Poisson}
  P_{\bar{\rm P}}(s)=\frac{\pi}{2}\,s\,e^{-\pi s^2/4}\:.
\end{equation}
This should not be confused with the Wigner distribution~(\ref{wigner}).

We report on simulations done with gauge group SU(2) on a $6^4$ lattice
using $\beta=4/g^2=1.3$ in the confinement region for $N_f=2$ flavors
of staggered fermions of mass $ma=0.07$. For this system the fermion
determinant is real and lattice simulations become feasible.\cite{Hand99}
We sampled
160 independent configurations.  In the case of color-SU(2), the
staggered Dirac operator has an extra anti-unitary symmetry
\cite{Hands90} and falls in the symmetry class with Dyson parameter
$\beta_D=4$.\cite{Hala97b} However, one can show that the
nearest-neighbor spacing distribution in the bulk of the spectrum is
also given by Eq.~(\ref{Ginibre}).

\begin{figure}
\begin{center}
\begin{tabular}{ccccc}
  & {\large $\mu=0$}  & \hspace*{6.5mm}   & &
  {\large $\mu=0.4$} \\
  \vspace*{0mm}
  & $ma=0.07$ &&& $ma=0.07$ \\[2mm]
  \multicolumn{2}{c}{\epsfxsize=5cm\epsffile{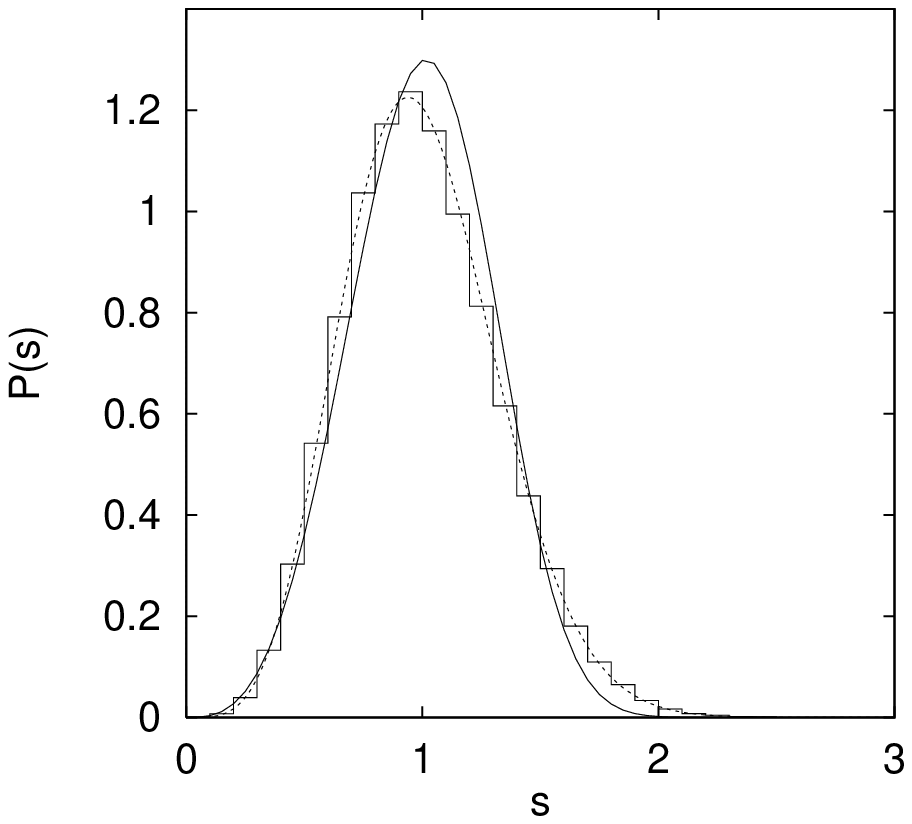}} &&
  \multicolumn{2}{c}{\epsfxsize=5cm\epsffile{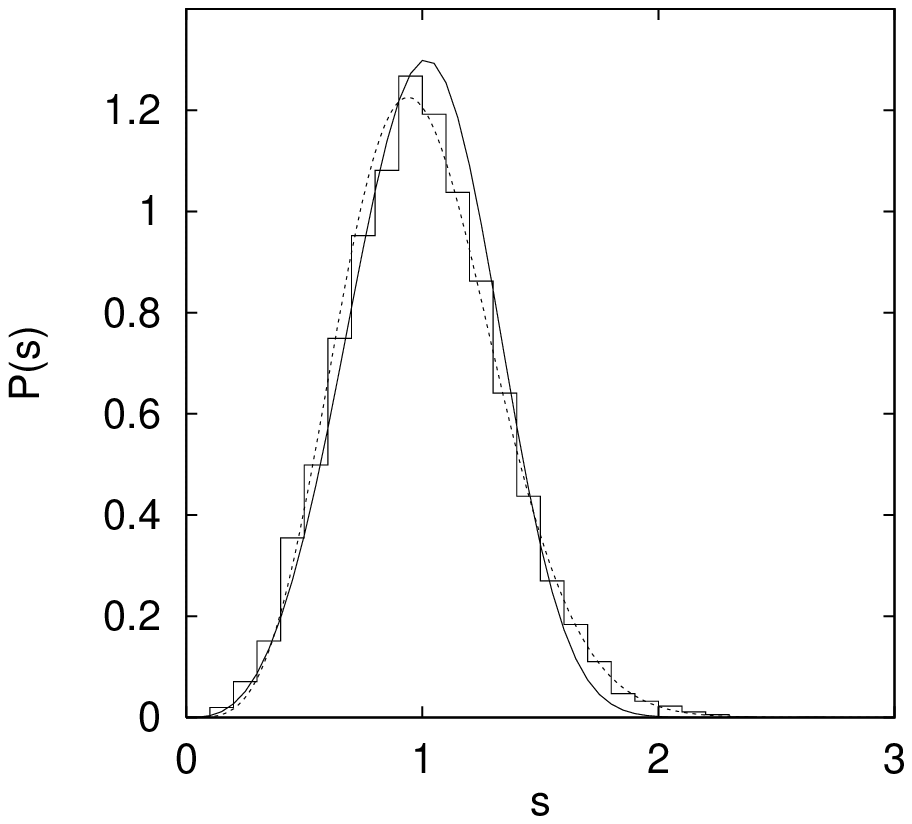}}
\end{tabular}
\end{center}
\vspace*{-3mm}
  \caption{Nearest-neighbor spacing distribution in the complex plane
    for two-color QCD with 
    $\mu$ in the confinement (left) and deconfinement (right) phase.
    The short-dashed curve is the Wigner distribution
    for the chSE and the solid curve is the Ginibre distribution of
    Eq.~(\protect\ref{Ginibre}).}
  \vspace*{-3mm}
  \label{Psconf}
\end{figure}

Our results for $P(s)$ are presented in Fig.~\ref{Psconf}.  As a
function of $\mu$, we expect to find a transition from Wigner to
Ginibre behavior in $P(s)$. This was clearly seen in color-SU(3) with
$N_f=3$ flavors and quenched chemical potential,~\cite{Mark99} where
differences between both curves are more pronounced. For the
symplectic ensemble of color-SU(2) with staggered fermions, the Wigner
and Ginibre distributions are very close to each other and thus harder
to distinguish. They are reproduced by our preliminary data for
$\mu=0$ and $\mu=0.4$, respectively.

For $\mu > 1.0$, the lattice results for $P(s)$ deviate substantially
from the Ginibre distribution and could be interpreted as Poisson
behavior, corresponding to uncorrelated eigenvalues. (In the Hermitian
case at nonzero temperature, lattice simulations only show a
transition to Poisson behavior for $\beta\to\infty$ when the physical
box size shrinks and the theory becomes free.\cite{Pull98}) A plausible
explanation of the transition to Poisson behavior is provided by the
following two (related) observations. First, for large $\mu$ the terms
containing $e^\mu$ in Eq.~(\ref{Dirac}) dominate the Dirac matrix, giving
rise to uncorrelated eigenvalues. Second, for large $\mu$ the fermion density
on the finite lattice reaches saturation due to the limited box size and
the Pauli exclusion principle.


\section{Low-lying Spectrum}

We have continued our investigations with a study of the
distribution of the small eigenvalues in the confined phase. The
Banks-Casher formula \cite{Bank80} relates the Dirac eigenvalue density
$\rho(\lambda)$ at $\lambda=0$ to the chiral condensate,
$ \Sigma \equiv |\langle \bar{\psi} \psi \rangle| =
 \lim_{\varepsilon\to 0}\lim_{V\to\infty} \pi\rho (\varepsilon)/V$.
The microscopic spectral density,
$ \rho_s (z) = \lim_{V\to\infty}
 \rho \left( {z/V\Sigma } \right)/V\Sigma , $
should be given by the appropriate result of RMT~\cite{ShVe92}, which
also generates the Leutwyler-Smilga sum rules.\cite{LeSm92}

\begin{figure}[t]
  \begin{center}
    \psfig{figure=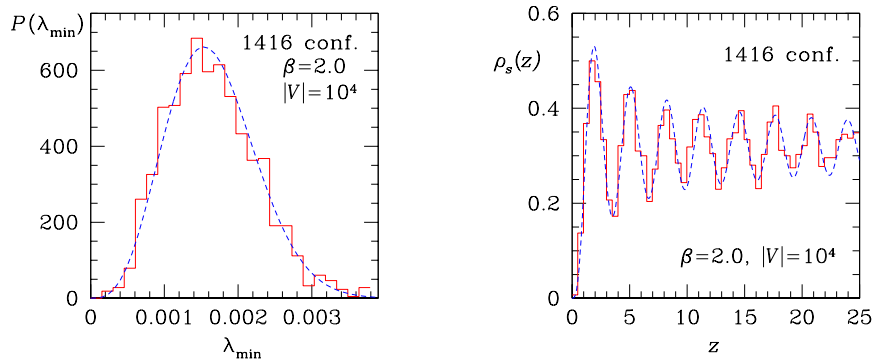,width=11.5cm}
  \end{center}
  \caption{Distribution $P(\lambda_{\min})$ (left) and microscopic
    spectral density $\rho_s (z)$ (right) from
    Ref.~\protect\onlinecite{Berb98a}  
    for SU(2) gauge theory on a $10^4$ lattice in comparison with the
    predictions of the chSE of RMT (dashed lines).}
  \label{f06a}
\end{figure}

To study the smallest eigenvalues, spectral averaging is not possible,
and one has to produce large numbers of configurations. 
We present results from Ref.~\onlinecite{Berb98a} for SU(2) theory and the
staggered Dirac operator. Both the distribution $P(\lambda_{\rm min})$ of the
smallest eigenvalue and the microscopic spectral density $\rho_s(z)$
agree with the RMT predictions of the chSE
for topological charge $\nu = 0$, as depicted in Fig.~\ref{f06a}.

\begin{figure}[!b]
  \begin{center}
    \hspace*{10mm}\psfig{figure=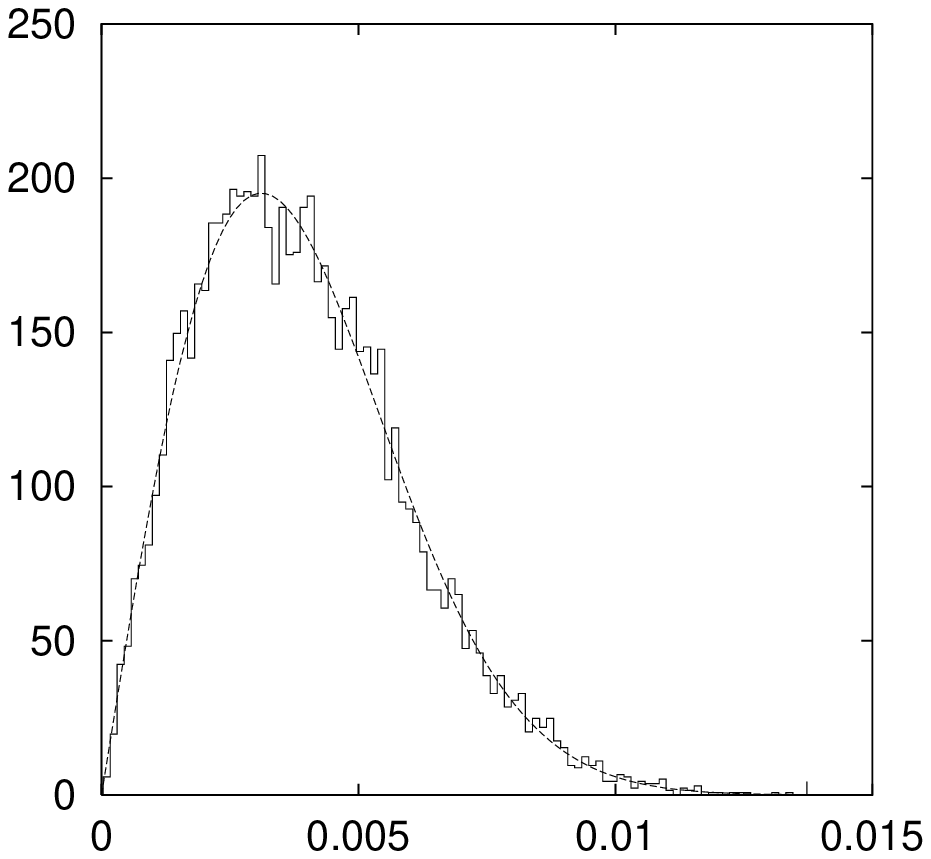,height=4cm}
    \hspace*{15mm}\psfig{figure=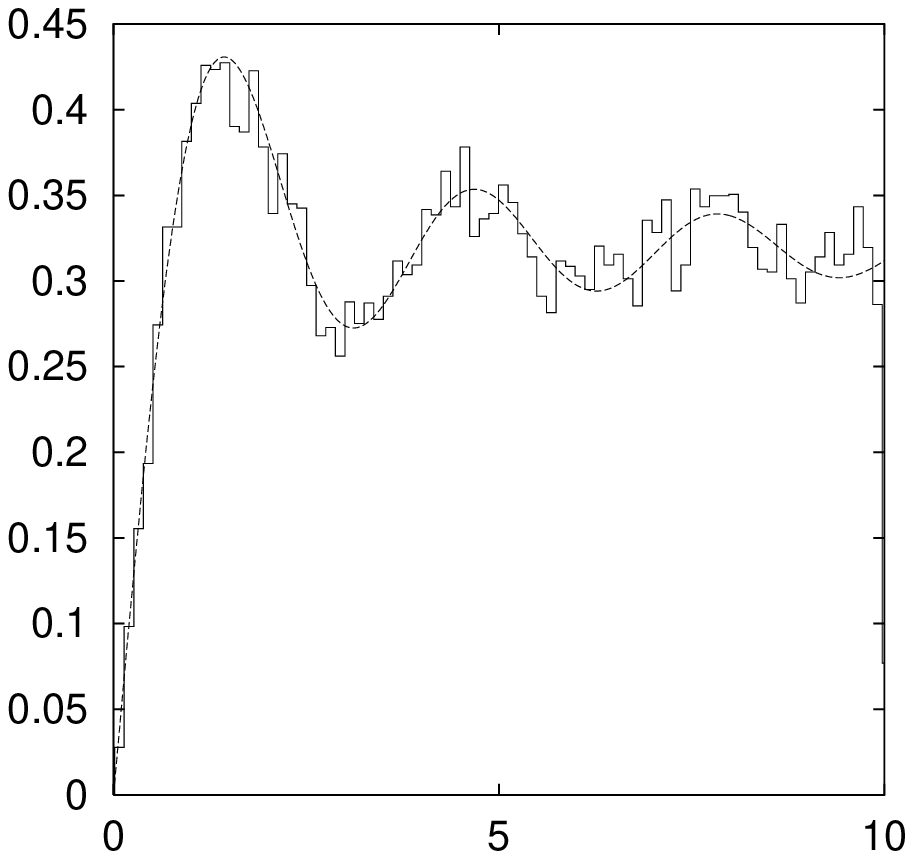,height=4cm}\\[-39.5mm]
    \hspace*{-45mm}$P(\lambda_{\min})$\hspace*{50mm}$\rho_s(z)$\\[35mm]
    \hspace*{38mm}$\lambda_{\min}$\hspace*{60mm}$z$
  \end{center}
  \caption{Distribution $P(\lambda_{\min})$ (left) and microscopic
    spectral density $\rho_s (z)$ (right) from our $6^4$ lattice data
    of U(1) gauge theory in comparison with the predictions of the chUE
    of RMT (dashed lines).}
  \label{f06}
\end{figure}

Our analog results for U(1) theory are for $\beta=0.9$ in the confined
phase with  10000 configurations on a $6^4$ lattice. The left plot in
Fig.~\ref{f06} 
exhibits the distribution $P(\lambda_{\min})$ of the
smallest eigenvalue in comparison with the prediction
of the (quenched) chUE of RMT for topological charge $\nu=0$,
$P(\lambda_{\min}) = (V\Sigma)^2 (\lambda_{\min}/2)\,\exp( - 
(V\Sigma\lambda_{\min})^2/4)$.
For the chiral
condensate we obtain $\Sigma \approx 0.35$ by extrapolating the
histogram for $\rho(\lambda)$ to $\lambda=0$ and using the
Banks-Casher relation. 
In the right plot of
Fig.~\ref{f06} the same comparison with RMT is done for the microscopic
spectral density $\rho_s (z)$ up to $z=10$, and the agreement is
again quite satisfactory. Here, the analytical RMT result for the
(quenched) chUE and $\nu=0$ is given by \cite{ShVe92}
$ \rho_s(z) = z\, [ J_0^2(z) + J_1^2(z) ]/2$, where $J$ denotes the
Bessel function.

\begin{figure}
  \begin{center}
    \psfig{figure=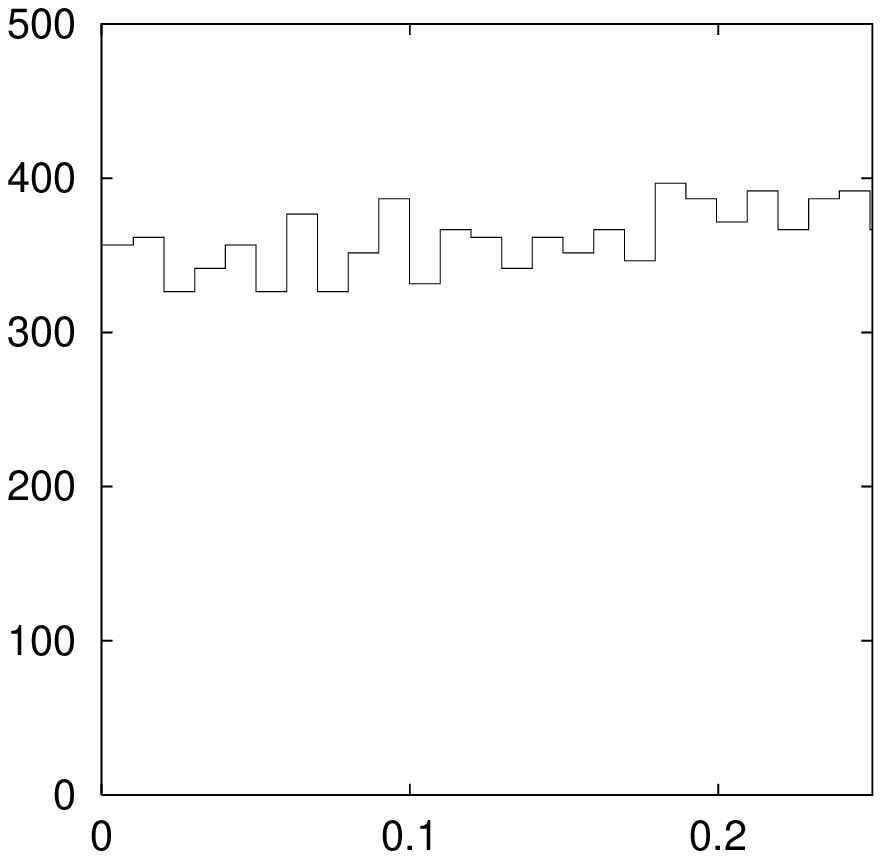,width=4cm}\hspace*{15mm}
    \psfig{figure=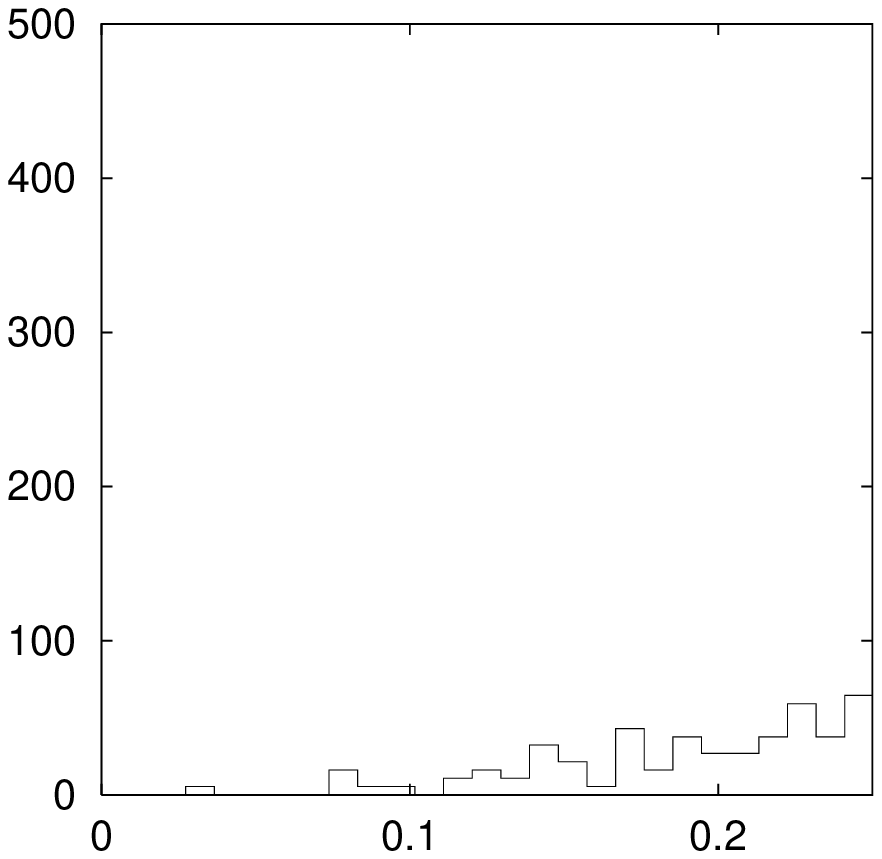,width=4cm}\\[-40mm]
    \hspace*{-50mm}$\rho(\lambda)$\hspace*{50mm}$\rho(\lambda)$\\[32mm]
    \hspace*{38mm}$\lambda$\hspace*{55mm}$\lambda$
  \end{center}
  \caption{Density $\rho(\lambda)$ of small eigenvalues for U(1) on an
    $8^3\times 6$ lattice at $\beta = 0.9$ (left) and at $\beta =
    1.1$ (right). A nonzero chiral condensate is supported in
    the confinement phase of U(1) gauge theory.}
  \label{f12}
\end{figure}

In U(1) theory, the quasi-zero modes which are responsible for
the chiral condensate
$\Sigma \approx 0.35$ build up when we cross from the Coulomb into the
confined phase. For our $8^3\times 6$ lattice, Fig.~\ref{f12} compares
on identical scales densities of the small eigenvalues at $\beta = 0.9
$ and at $\beta = 1.1$, averaged over 20 configurations.
The quasi-zero modes in the left plot are related to the nonzero
chiral condensate, whereas no
such quasi-zero modes are found in the Coulomb phase. 
It would be worthwhile to understand the physical
origin of the U(1) quasi-zero modes in more detail. For 4d SU(2) and
SU(3) gauge theories a general interpretation is to link them, and
hence the chiral condensate, to the existence of instantons.  In 4d
U(1) gauge theory it has been shown numerically that the chiral
condensate is mainly due to monopole configurations.\cite{BHSW98}

\begin{figure}
   \begin{center}
     \psfig{figure=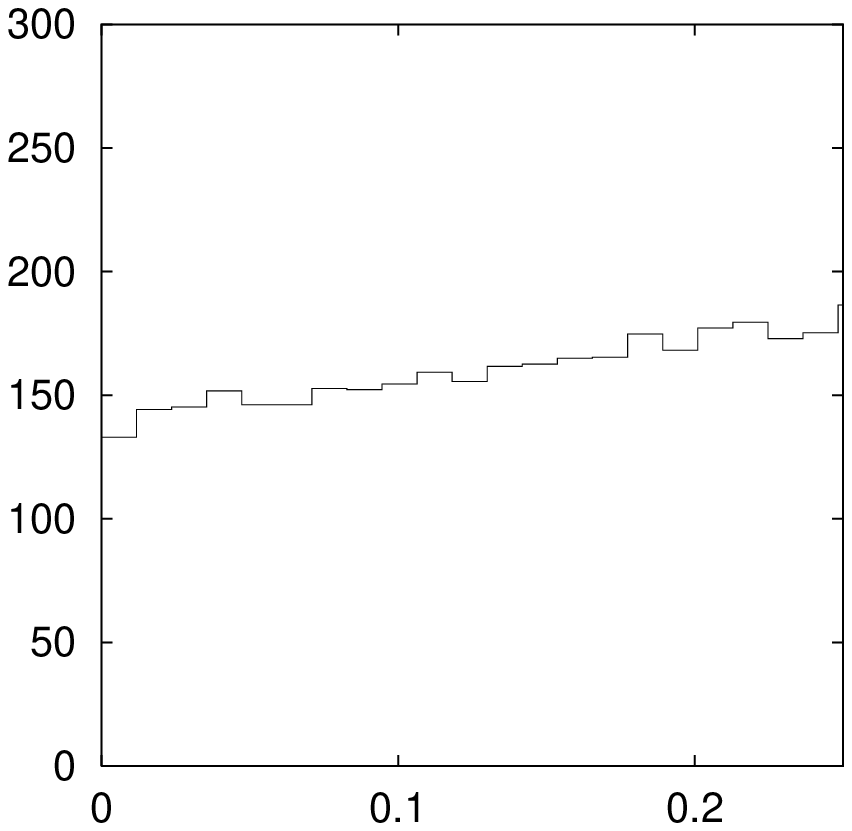,width=4cm}\hspace*{15mm}
     \psfig{figure=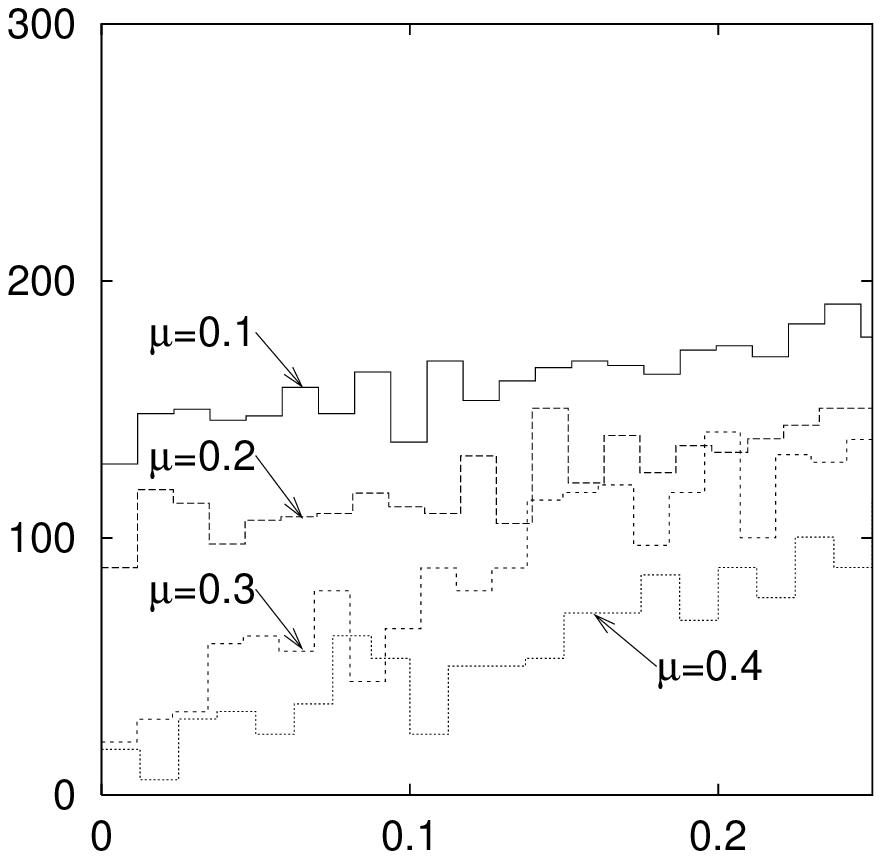,width=4cm}\\[-40mm]
     \hspace*{-50mm}$\rho(y)$\hspace*{50mm}$\rho(y)$\\[32mm]
     \hspace*{38mm}$y$\hspace*{55mm}$y$
   \end{center}
   \caption{Density $\rho(y)$ of small eigenvalues (see text)
     for two-color QCD 
     on a $6^4$ lattice at $\mu=0$ (left) and at $\mu=0.1$ to
     0.4 (right). 
     The vanishing of the chiral condensate in the matter phase, found in
     Ref.~\protect\onlinecite{Hand99} at $\mu_c\approx0.3$, is
     accompanied by a loss of quasi-zero modes.}
   \label{f12a}
 \end{figure}

For two-color QCD, numerical simulations of the full theory with
chemical potential in Ref.~\onlinecite{Hand99} have exhibited a chiral
phase
transition at $\mu_c \approx 0.3$ where the chiral condensate as the
associated order parameter vanishes. In Fig.~\ref{f12a} we compare the
densities of the small eigenvalues at $\mu=0$ and at $\mu=0.1$ to 0.4 on our
$6^4$ lattice, averaged over 160 configurations. Since the eigenvalues
move into the complex plane for $\mu > 0$, a band of width $\epsilon
= 0.015$ parallel to the imaginary axis is considered to construct
$\rho(y)$, i.e. $\rho(y)\equiv\int_{-\epsilon}^\epsilon
dx\,\rho(x,y)$, where $\rho(x,y)$ is the density of the complex
eigenvalues $x+iy$. 

\begin{figure}
  \begin{center}
    \psfig{figure=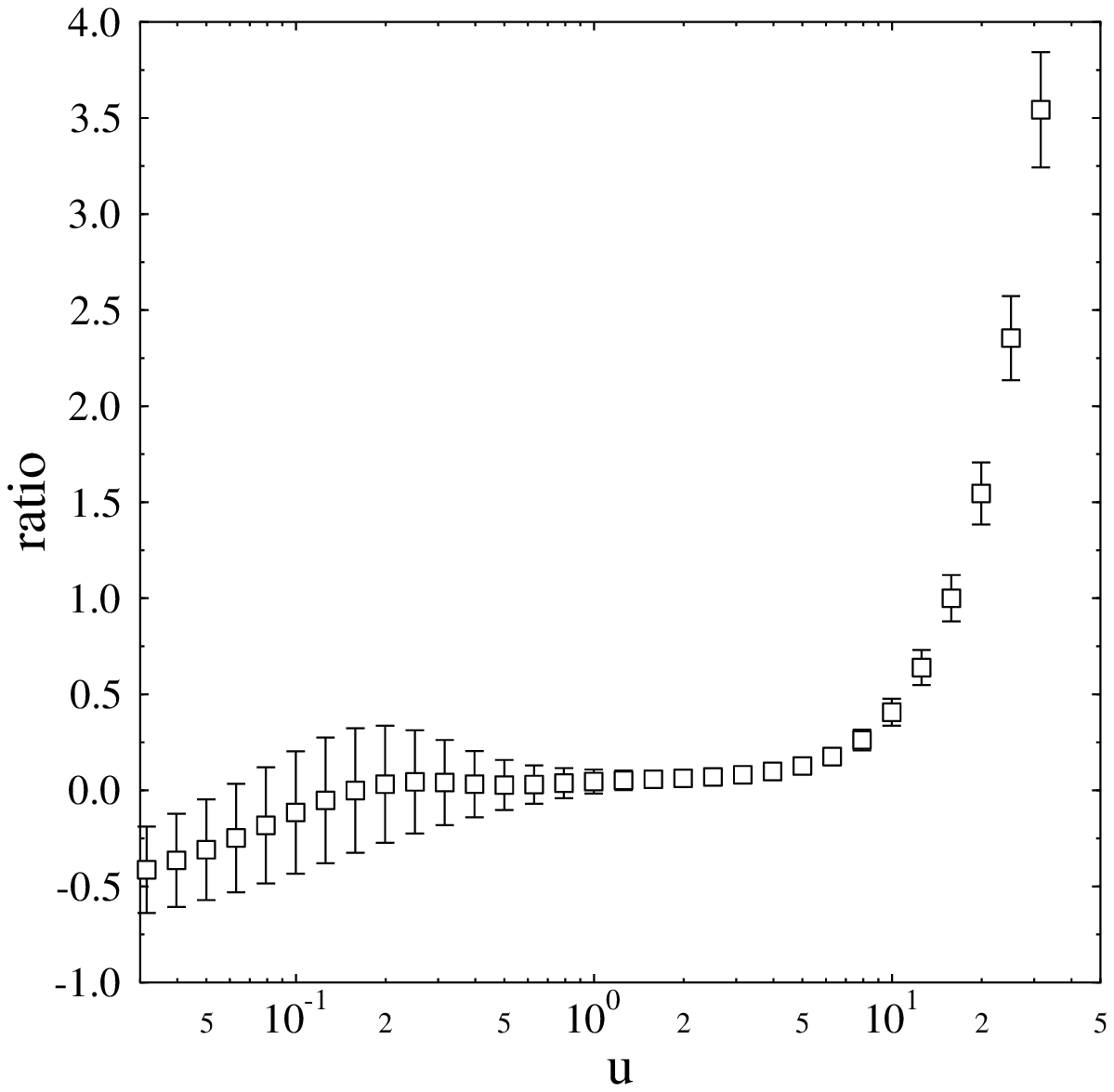,height=50mm}\hspace*{10mm}
    \psfig{figure=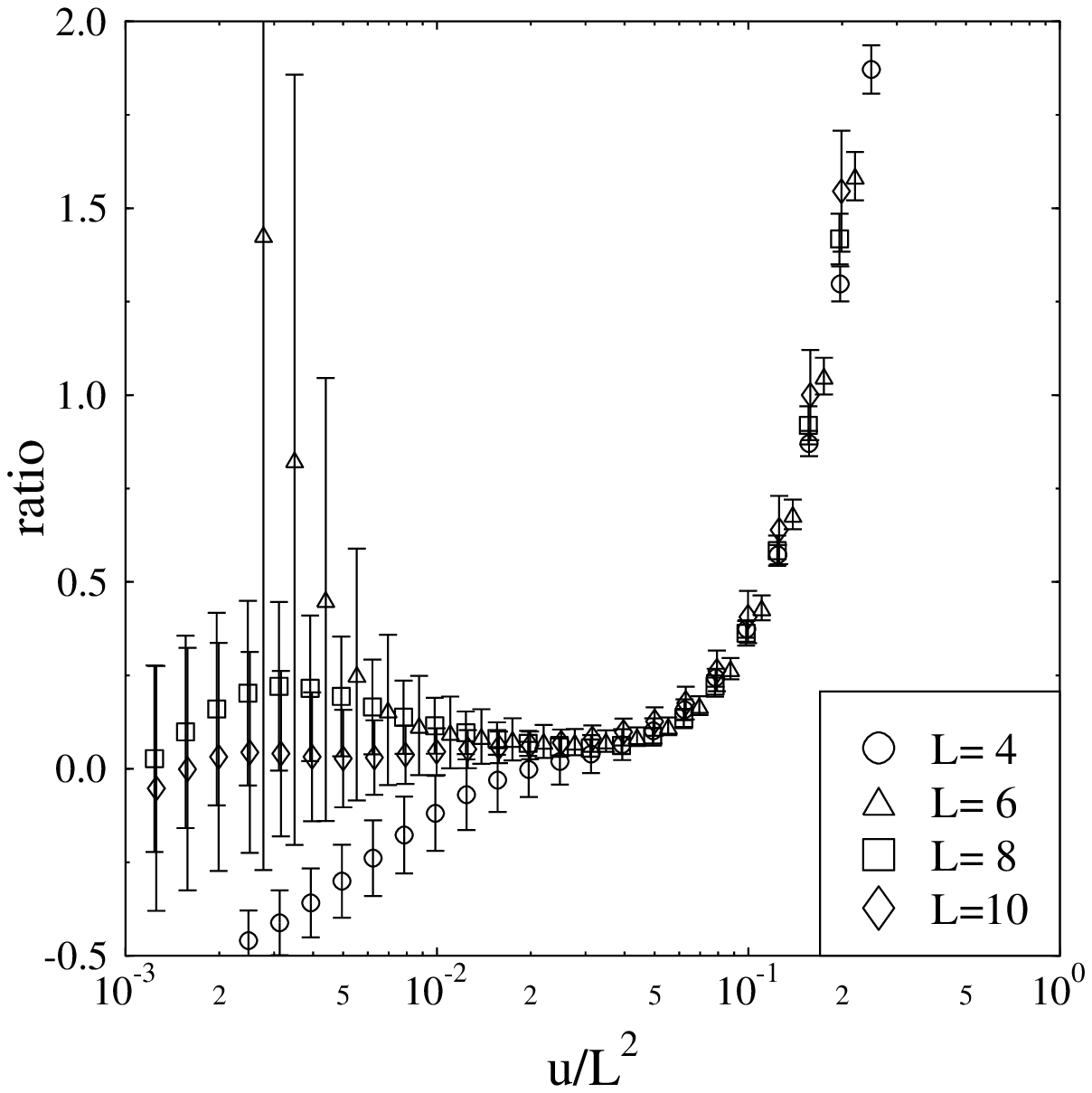,height=50mm}
    \caption{The ratio of Eq.~(\ref{ratio}) for SU(2) gauge theory
      plotted versus $u$ and $u/(L_sL_t)$, respectively (in this case,
      $L_s=L_t=L$).  The left plot is for $L=10$.  In the right plot,
      the data for different $L$ fall on the
      same curve, confirming the expected scaling of the Thouless
      energy according to Eq.~(\ref{thouless}).  The deviations of the
      ratio from zero for very small values of $u$ are well-understood
      artifacts of the finite lattice size and of finite statistics.
      \protect\cite{Berb98}}
    \label{fig8a}
  \end{center}
\end{figure}

\begin{figure}
  \begin{center}
    \psfig{figure=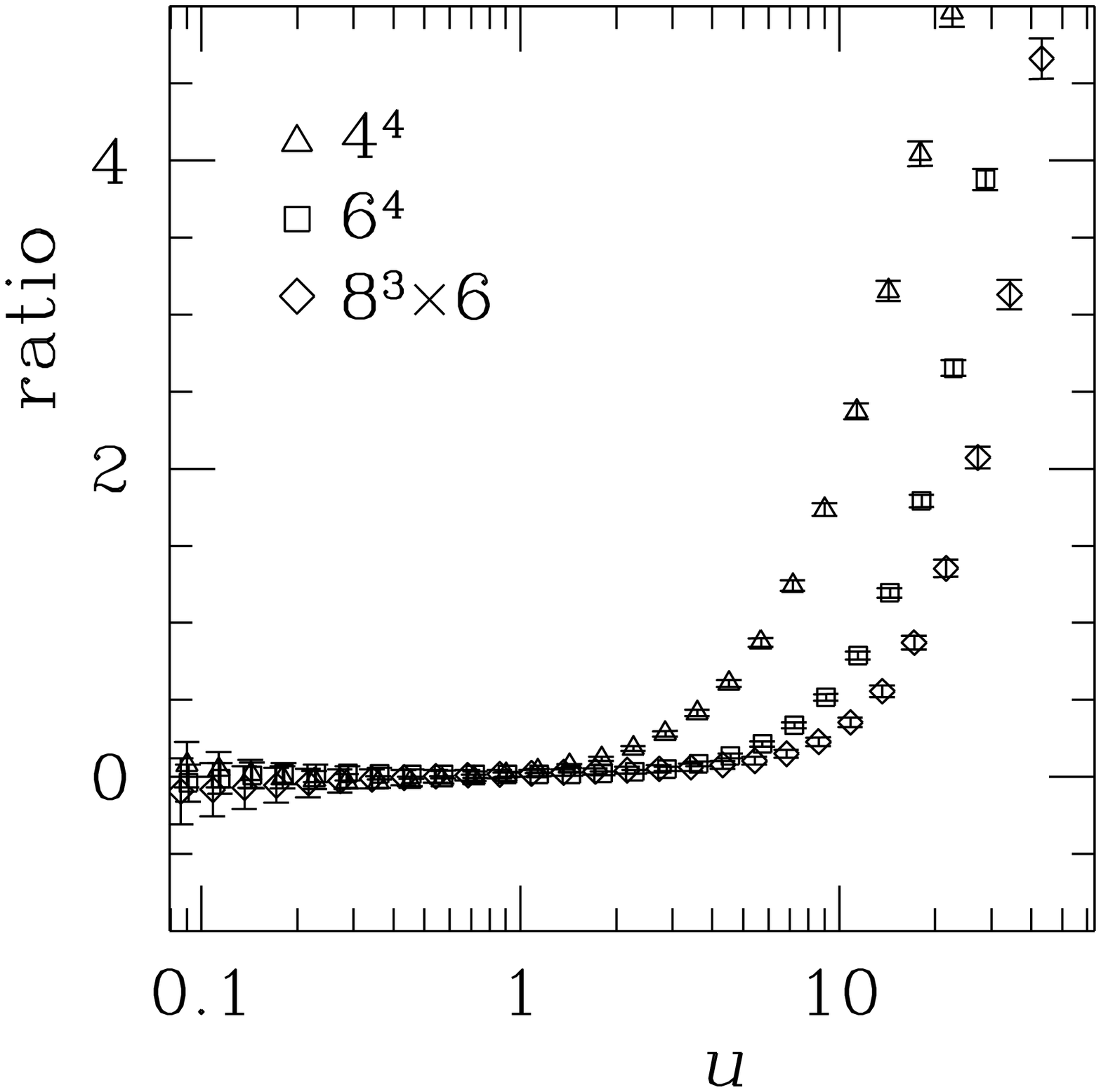,height=50mm}\hspace*{10.3mm}
    \psfig{figure=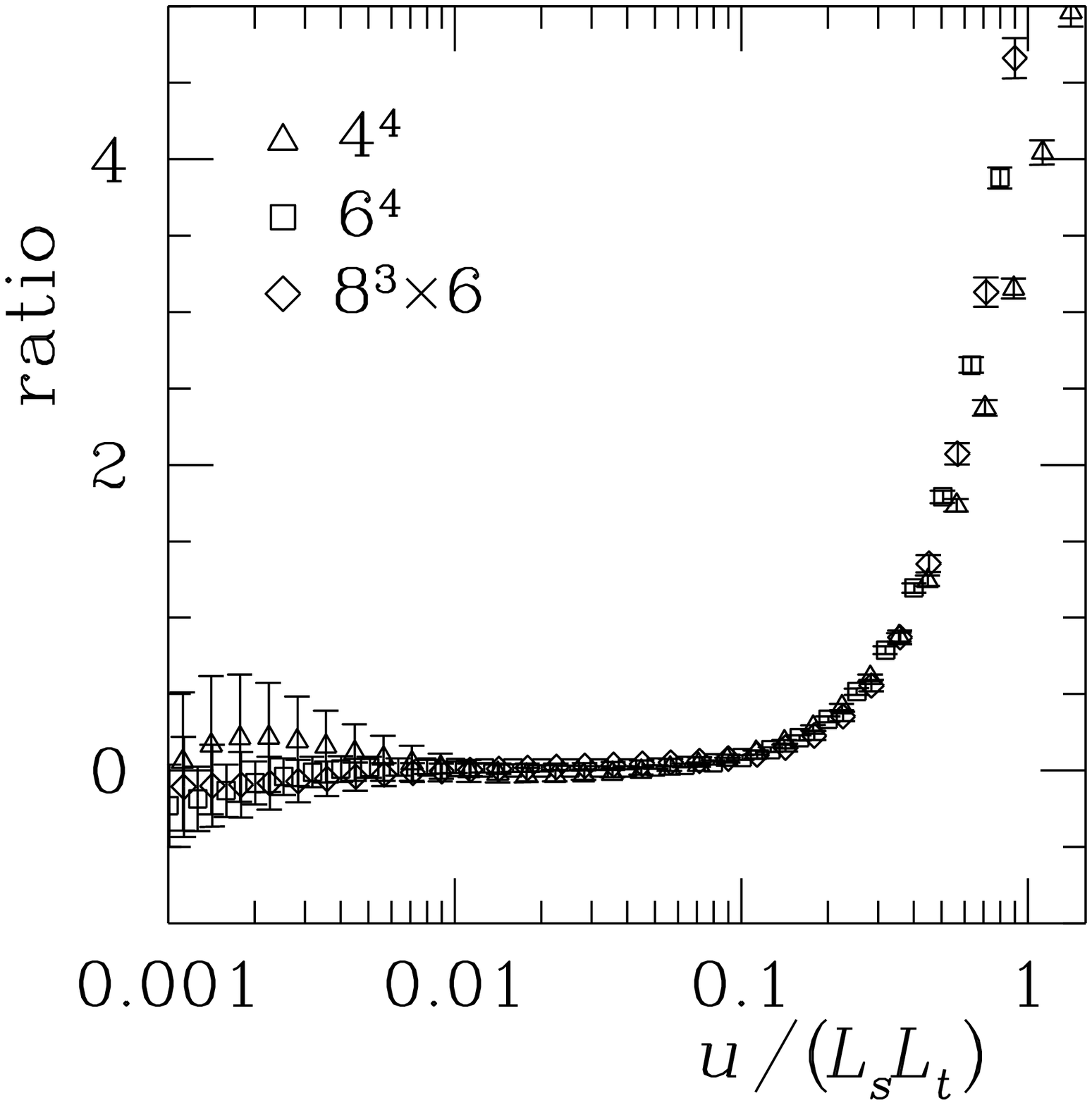,height=50mm}
    \caption{Same as Fig.~\ref{fig8a} but for U(1) gauge theory.}
    \label{fig8}
  \end{center}
\end{figure}

Another interesting question concerns the energy scale $E_c$ up to
which RMT describes the small Dirac eigenvalues in the phase where
$\Sigma>0$.  In disordered mesoscopic systems, a similar scale is
called the Thouless energy.  The theoretical prediction for QCD is
$E_c\sim f_\pi^2/\Sigma L^{2}_{s}$ (see Ref.~\onlinecite{Osbo98})
with the pion decay
constant $f_{\pi}$, where we have assumed that the spatial extent
$L_s$ of the lattice is not smaller than the temporal extent $L_t$.
In units of the mean level spacing $\Delta=\pi/V\Sigma$ at the origin,
this becomes
\begin{equation}
  \label{thouless}
  u_c\equiv \frac{E_c}{\Delta} \sim \frac1\pi f_\pi^2L_sL_t\:.
\end{equation}
A convenient quantity from
which $u_c$ can be extracted is the disconnected scalar
susceptibility,
\begin{equation}
  \chi_{\rm latt}^{\rm disc}(m)=\frac1N\left\langle\sum_{k,l=1}^N
    \frac1{(i\lambda_k+m)(i\lambda_l+m)}\right\rangle_{\!\!A}
  -\frac1N\left\langle
    \sum_{k=1}^N\frac1{i\lambda_k+m}\right\rangle_{\!\!A}^2\:.
\end{equation}
The corresponding RMT result for the quenched chSE with $\nu=0$
is given by~\cite{Berb98} $  \chi^{\rm disc}_{\rm RMT}
= 4u^2 \int_0^1 ds\: s^2K_0(2su) \int_0^1 dt\:
  I_0(2stu)\{s(1-t^2)
  +4K_0(2u)[I_0(2su) + $ $tI_0(2stu)]
  -8stI_0(2stu)K_0(2su)\}
  -4u^2K_0^2(2u) [ \int_0^1 ds \:I_0(2su)]^2$,
and for the quenched chUE with $\nu=0$ we have
\cite{Goec99} $\chi_{\rm RMT}^{\rm
  disc}=u^2[K_1^2(u)-K_0^2(u)][I_0^2(u)-I_1^2(u)]$.  Here,
$u=mV\Sigma$, and $I$ and $K$ are modified Bessel functions.  In
Figs.~\ref{fig8a} and \ref{fig8} we have plotted the ratio \cite{Berb98}
\begin{equation}
  \label{ratio}
  {\rm ratio}=\left(\chi_{\rm latt}^{\rm disc}-\chi_{\rm RMT}^{\rm
      disc}\right)/\chi_{\rm RMT}^{\rm disc}
\end{equation}
versus $u$ and $u/(L_sL_t)$, for color-SU(2) with staggered Dirac
operator at $\beta=2.0$ from Ref.~\onlinecite{Berb98} and for the U(1)
data computed 
at $\beta=0.9$, respectively.  This ratio should deviate from zero above the
Thouless scale.  The expected scaling of the Thouless energy with
$L_sL_t$ is confirmed.


\section{Conclusions}

The aim of this paper was to work out two different types of
universalities inherent in quantum field theories and their
interpretation in terms of RMT. The first type concerns the bulk of
the spectrum of the Dirac operator.  The nearest-neighbor spacing
distribution $P(s)$ agrees with the RMT result in both the confinement
and the deconfinement phase of pure gauge theory and of full QCD,
except for extremely large values of $\beta$ where the eigenvalues are
known analytically.  
The nearest-neighbor spacing distribution of 4d U(1) quenched lattice
gauge theory is described by the chUE of RMT in both the confinement
and the Coulomb phase.
Even in the Coulomb as well as in the deconfinement phase,
gauge fields retain a considerable degree of randomness, which 
apparently gives rise to quantum chaos in these theories.

A general unfolding procedure for the spectra of non-Hermitian operators
was applied to the lattice Dirac operator for two-color QCD at nonzero
chemical potential.
Agreement of the nearest-neighbor spacing distribution in the complex
plane with predictions of
the Ginibre ensemble of non-Hermitian RMT was found around $\mu=0.4$.
The changes for larger values of $\mu$ toward a Poisson distribution
are understood formally. The physical interpretation requires a better 
understanding of QCD at nonzero density. 

The second type of universality concerns the low-lying spectra of the
Dirac operators of both QCD and QED. In all cases considered, one
finds that in the phase in which  chiral symmetry is spontaneously
broken the distribution
$P(\lambda_{\min})$ and the microscopic spectral density $\rho_s(z)$
are described by RMT.
The Thouless energy scales with the lattice size as expected.

In summary, both the bulk and the low-lying Dirac spectrum are
governed by RMT and the related symmetries.


\section{Acknowledgments}

This study was supported in part by FWF project P11456-PHY, by DOE
contracts DE-FG02-97ER41022, DE-FG05-85ER2500, DE-FG02-91ER40608, and
DE-AC02-98CH10886, and by the RIKEN BNL Research Center.


\end{document}